\documentclass[a4,conference]{IEEEtran}
\ifCLASSINFOpdf
\else
\fi
\usepackage{graphicx}
\usepackage{color}
\usepackage{epstopdf}
\usepackage{setspace}
\usepackage{cite}
\usepackage{amsmath}
\usepackage{amssymb}
\newtheorem{lemma}{Lemma}
\newtheorem{remark}{Remark}
\DeclareMathSizes{10}{9}{7}{6}
\IEEEaftertitletext{\vspace{-3\baselineskip}}
\begin{document}
\title{Dynamic HARQ with Guaranteed Delay}
\author{Mahyar~Shirvanimoghaddam\IEEEauthorrefmark{1}, Hossein Khayami\IEEEauthorrefmark{2}, Yonghui Li\IEEEauthorrefmark{1}, Branka Vucetic\IEEEauthorrefmark{1}\\
\IEEEauthorrefmark{1}School of Electrical and Information Engineering, The University of Sydney, NSW, Australia\\
\IEEEauthorrefmark{2} School of Electrical Engineering, Sharif University of Technology, Tehran, Iran\\
Emails: \{mahyar.shm, yonghui.li, branka.vucetic\}@sydney.edu.au, h.khayami@alum.sharif.edu}
\maketitle
\begin{abstract}
In this paper, a dynamic-hybrid automatic repeat request (D-HARQ) scheme with guaranteed delay performance is proposed. As opposed to the conventional HARQ that the maximum number of re-transmissions, $L$, is fixed, in the proposed scheme packets can be re-transmitted more times given that the previous packet was received with less than $L$ re-transmissions. The dynamic of the proposed scheme is analyzed using the Markov model. For delay sensitive applications, the proposed scheme shows a superior performance in terms of packet error rate compared with the conventional HARQ and Fixed re-transmission schemes when the channel state information is not available at the transmitter. We further show that D-HARQ achieves a higher throughput compared with the conventional HARQ and fixed re-transmission schemes under the same reliability constraint. 
\end{abstract}
\IEEEpeerreviewmaketitle

\section{Introduction}
The 3$^\text{rd}$ generation partnership project (3GPP) has identified ultra-reliable low-latency communications (URLLC) as one of the key application scenarios for the 5$^\text{th}$ generation (5G) mobile standard \cite{chen2018ultra}. URLLC mostly covers mission critical applications, such as smart grids and wireless industrial control. URLLC demands a very high reliability and low latency and targets transmissions of small data payloads and accordingly short packets to meet the latency requirement \cite{ostman2018low}. The general URLLC requirement according to 3GPP is that the packet error rate (PER) for the transmission of packets of size 32 bytes should be $\le 10^{-5}$, within a user plane latency of 1ms (with or without re-transmission) \cite{TS-22.261}. The user plane latency is defined as the time to successfully deliver a data block from the transmitter to the receiver via the radio interface in both uplink and downlink directions \cite{Shirvani2019Short}.

To meet the latency constraint in URLLC applications one can widen the system bandwidth, which is not always possible especially for applications that might operate over unlicensed spectrum \cite{Zhibo2018WirelessHP}. We can also use short packet communications to reduce the latency but this will reduce the coding gain \cite{Shirvani2019Short}. This is mainly because of the fewer channel observations due to the finite length. More specifically, if we decrease the block length, the coding gain will be reduced and the gap to the Shannon's limit will increase \cite{polyanskiy2010channel}. Authors in \cite{polyanskiy2010channel} characterized the channel coding rate in the finite length regime. For short packet communications in URLLC, we need to use strong channel codes paired with re-transmission techniques for enhancing reliability, which indeed increase the latency \cite{bennis2018ultrareliable}. 

In most re-transmission strategies, a packet is re-transmitted several times when the decoding failed. The receiver sends a negative acknowledgment (NACK) to the transmitter and asks for a re-transmission. In automatic repeat request (ARQ), the receiver drops all the previous copies of the packet and performs the decoding on the freshly arrived copy of the packet. In Hybrid ARQ (HARQ), the receiver keeps all the previous copies and combines them with the freshly received packet for the decoding. The re-transmission stops when the a pre-determined number of copies are sent and the receiver sent an acknowledgment (ACK). In Chase combining (CC) HARQ, the transmitter sends the same packet every time it receives a NACK and the receiver uses maximum ratio combining (MRC) to combine packets and performs the decoding. The transmitter can also generate new encoded packets and send them to the receiver, so the receiver can perform the decoding on a longer codeword. This is referred to as incremental redundancy (IR). Both CC-HARQ and IR-HARQ have been studied under the finite block length assumption in the additive white Gaussian noise (AWGN) channel \cite{kim2013performance}, Rayleigh block fading channel \cite{wu2011coding,lee2015harq}, and other communication scenarios \cite{makki2014finite,avranas2018energy}. 

Authors in \cite{ostman2018low}, considered the low latency short packet communications and evaluated the HARQ and fixed transmission strategy. In fixed transmission (Fixed-Tx) each packet will be re-transmitted several times and the number of re-transmissions is set in advance. The authors also considered the delay associated with the feedback message in the HARQ scheme. As shown in \cite{ostman2018low}, HARQ may significantly outperform Fixed-Tx in terms of the achievable rate under the same latency and reliability constraints for a given information block length and number of diversity branches. Motivated by this work, we propose a dynamic HARQ (D-HARQ) scheme, which targets ultra-reliable communications for delay sensitive applications. 

In D-HARQ, each packet can be re-transmitted more times given that the previous packet was decoded earlier than its deadline. In other words, in D-HARQ the diversity branches are opportunistically used by the transmitter in order to provide a higher level of reliability. We analyze the packet error rate and throughput of the proposed scheme and show that it outperforms both the Fixed-Tx and conventional HARQ schemes in terms of reliability and throughput under the same delay constraint. We particularly show that when only one re-transmission is allowed, which is the scenario of interest in many URLLC application \cite{rao2018packet}, the proposed D-HARQ scheme achieves more than 10dB gain compared with the conventional HARQ scheme under the same packet error rate of $10^{-3}$ and delay constraint. The gain further increases when higher levels of reliability are of interest and more diversity branches are available. 

The rest of the paper is organized as follows. In Section II, the system model and some preliminaries on HARQ and finite block length approximation are presented. In Section III, we explain the proposed dynamic HARQ scheme. We analyze the performance of D-HARQ in terms of packet error rate and throughput in Section IV.  Numerical results are represented in Section V. Finally, Section VI concludes the paper.

\section{System Model and Preliminaries}
\subsection{Channel Model}
We consider a memoryless Rayleigh block-fading channel represented by
\begin{align}
    y(t)=h(t)x(t)+w(t),
\end{align}
where $x(t)$ and $y(t)$ are the transmitted and received signals, respectively, $w(t)\sim \mathcal{CN}(0,1)$ is the zero-mean additive white Gaussian noise (AWGN) with power spectral density $N_0/2$, $h(t)\sim\mathcal{CN}(0,1)$ is the multipath fading component. We further assume that $E[|x(t)|^2]=P$ and $E[|h(t)|^2]=1$. The channel signal-to-noise ratio (SNR) is then given by $\gamma_0=P$. The channel stays constant within the coherence block and changes independently across coherence blocks. The channel coherence time and bandwidth are denoted by $T_c$ and $B_c$, respectively. For a system bandwidth $B$, the available number of diversity branches is $L_c=\lfloor B/B_c\rfloor$. 

\begin{figure}[t]
\centering
\includegraphics[width=1\columnwidth]{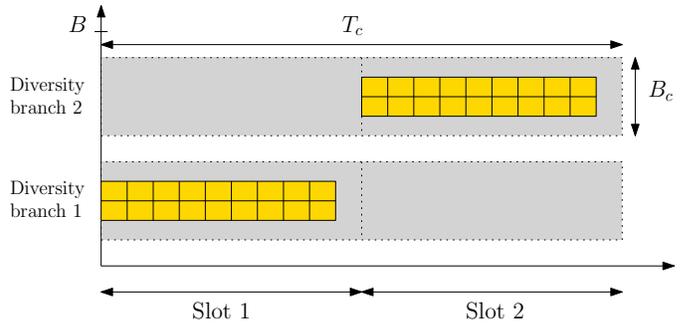}
\vspace{-2ex}
\caption{The block-fading OFDM channel model with $L_c=2$ diversity branches where $T_c$ spans 2 timeslots.}
\label{fig:systemmodel}
\end{figure}

For an orthogonal frequency division multiplexing (OFDM) system, each resource block (RB) consists of a number of OFDM symbols, each spanning a number of subcarriers. A time slot is defined as the interval over which an RB is transmitted. In each time slot, the transmitter sends over $L$ diversity branches. We assume that the transmission in different slots occur on different diversity branches and that $L_c\gg L$. Fig. \ref{fig:systemmodel} shows the block-fading OFDM channel model, where the number of diversity branches is $L_c=2$ and the channel coherence time $T_c$ spans 2 time slots. For simplicity of analysis in this paper, we assume that $L=1$ and packets experiences an independent fading over each time slot \cite{makki2019fast}. The extension to the general system setup considered in \cite{ostman2018low} is straightforward.

\subsection{Fixed Transmission and HARQ}
We consider a delay sensitive application that each packet needs to be received by a given deadline. More specifically, packet $\ell$ needs to be received on or before time $T_\ell$. We assume that the interval between $T_\ell$ and $T_{\ell-1}$ spans $L$ time slots for all packets. In Fixed-Tx, the transmission of packet $\ell$ starts at time $T_{\ell-1}$ and finishes at $T_{\ell}$. That is $L$ copies of the packet is received at the destination and the packet will be dropped if the decoding failed at $T_\ell$. Fig. \ref{fig:schemes}-(b) shows the Fixed-Tx scheme when $L=2$. Each packet is of length $2n$ symbols and that a rate $R$ channel code is used to encode $k$ information bits to $2n$ coded symbols, i.e., $R=k/(2n)$. 

In HARQ, each packet is sent by the transmitter and will be re-transmitted if a NACK is received by the transmitter. For simplicity, we assume that the transmitter re-transmits the exact same message in each round of re-transmissions and that the maximum number of re-transmissions is $L-1$. The re-transmission is stopped when the maximum number of re-transmissions is reached or an acknowledgment (ACK) is received by the transmitter. The receiver combines all copies of the same packet using maximum ratio combining (MRC) and performs maximum likelihood (ML) decoding. The packet is dropped when the maximum number of re-transmissions is reached and the receiver fails to decode the packet. We refer to this event as the \textit{packet failure}. 

In HARQ, the receiver needs to feedback an ACK/NACK on its data to inform the transmitter whether to re-transmit or terminate the current packet transmission. Such a feedback mechanism introduces a delay compared with the fixed transmission strategy. We assume that the feedback delay per time slot amount to half a timeslot. Therefore, using a HARQ strategy, a timeslot of length $2n$ symbols is further divided into two mini-slots, and the first mini-slot is for packet transmission and the second mini-slot is for the feedback message (see Fig. \ref{fig:schemes}-(a)). As each packet is sent over mini-slots of length $n$ symbols, as opposed to the timeslots of length $2n$ symbols in Fixed-TX, the channel code rate in HARQ is twice that in Fixed-TX. One can easily extend this model to consider an arbitrary length for the feedback message.

\subsection{Packet Failure Model}
Let us assume that $k$ information bits are sent using a packet of length $N$ symbols and re-transmitted $m-1$ times over $m$ independent diversity branches. For simplicity we assume that all packets, including the re-tranmitted packets, have the same length $N$. Using maximum ratio combining (MRC) at the receiver, the overall signal-to-noise ratio (SNR) is given by $\gamma_{CC}=\sum_{i=1}^m \gamma_i$, where $\gamma_i$ is the channel SNR at time slot $i$. We use the normal approximation \cite{polyanskiy2010channel} to find the packet error rate for CC-HARQ, which is given by
\begin{align}
\epsilon^{(\mathrm{CC})}_m&(k,N)\approx Q\left(\frac{N\log_2(1+\sum_{i=1}^m\gamma_i)-k\log_2(N)}{\sqrt{NV(\sum_{i=1}^m\gamma_i)}}\right),
    \label{eq:errorcc}
\end{align}
where $Q(x)=\frac{1}{\sqrt{2\pi}}\int_{x}^{\infty}e^{-\frac{u^2}{2}}du$ is the standard $Q$-function and $V(\gamma) = \frac{\gamma(\gamma+2)}{(\gamma+1)^2}\log^2_2(e)$ is the channel dispersion \cite{polyanskiy2010channel}. 

For IR-HARQ, the packet error rate can be approximated by \cite{sahin2019delay}:
\begin{align}
  \epsilon^{(\mathrm{IR})}_m&(k,N)\approx Q\left(\frac{N\sum_{i=1}^{m}\log_2(1+\gamma_i)-k\log_2(mN)}{\sqrt{N\sum_{i=1}^mV(\gamma_i)}}\right),
    \label{eq:errorir}
\end{align}
 where we assumed that the re-transmitted packets are of the same length $N$. One can easily extend this model to a more general case that each re-transmission is of a different length\footnote{The bound can be closely approached by using rateless codes which can be designed for a wide range of SNRs and allows for variable length packet transmission, see for example \cite{RanaShortAFC2019,SachiniRaptor2018}. }.




\section{Dynamic HARQ with Guaranteed Delay}
In conventional HARQ, the maximum number of re-transmissions, $L-1$, is limited and is set in advance. However, in the proposed scheme a packet can be re-transmitted more than $L-1$ times given that the previous packet was received earlier than its deadline. The number of additional retransmission is dynamic and depends on the decoding of the previous packet. In Dynamic HARQ (D-HARQ), each packet can be re-transmitted, through either Chase combining (CC) or incremental redundancy (IR), until the packet is received correctly and accordingly an acknowledgment is received at the transmitter, or the delay constraint is violated, therefore the packet is dropped. We show that D-HARQ is superior to conventional HARQ in terms of reliability while maintaining the same level of throughput.
\begin{figure}[t]
\centering
\includegraphics[width=0.9\columnwidth]{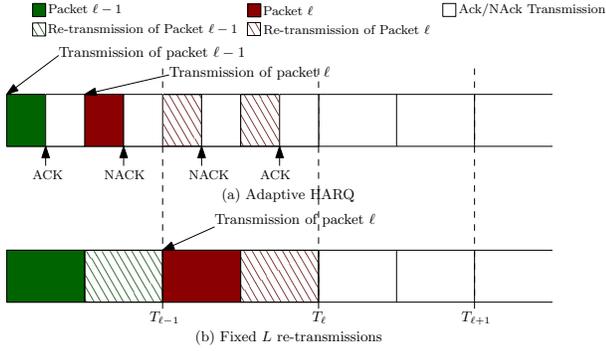}
\caption{Re-transmission techniques; (s) The proposed Adaptive HARQ and (b) Fixed re-transmission numbers. The feedback delay per transmission round amounts to one slot. }
\label{fig:schemes}
\end{figure}

As can be seen in Fig. \ref{fig:schemes}-(a), packet $\ell-1$ is correctly received one time slot earlier than its deadline, therefore the transmission of packet $\ell$ can be started earlier, leaving more transmission opportunities for packet $\ell$. This means that packet $\ell$ can be transmitted $L+1$ times via either Chase combining or incremental redundancy. Fig. \ref{fig:schemes}-b shows the fixed transmission strategy, where each packet is transmitted exactly $L$ times. As the transmitter does not need feedback, the transmitter can use longer packets for transmitting data. That is for a given data block of length $k$ bits, the transmitter in the Fixed-Tx scheme can use a rate $R=k/(2n)$ channel code, while in D-HARQ a rate $2R$ channel code should be used, assuming that the same modulation is being used. 

We limit the maximum number of transmissions in D-HARQ to $L+m$, for $0\le m<L$; that is the transmission of the packet it terminated when at most $L+m$ copies of the packet is sent or the deadline is reached or an acknowledgment is received, whichever is earlier. This constraint limit the energy consumption per packet of data, otherwise the number of re-transmissions can be potentially large when the packets are decoded much earlier than their deadlines, especially when the channel condition is good. It is important to note that when $m=0$, the D-HARQ will be equivalent to HARQ, as the packet can be transmitted $L$ times only, even if the previous packet was decoded earlier than its deadline. 



\section{Reliability and Throughput Analysis}
In this section, we evaluate the packet error rate of the proposed D-HARQ for a given packet deadline and time slot duration. We focus on short packet communications and use normal approximation \cite{polyanskiy2010channel} to calculate the error probability as in (\ref{eq:errorcc}) and (\ref{eq:errorir}). Let $\mathcal{A}_w$ denotes an event that the decoding failed for a packet of length $n$ symbols encoded with a rate $R=k/n$ channel code, which has been transmitted $w$ times, then $\mathrm{Prob}\{\mathcal{A}_w\}=\epsilon_{w}(k,n)$, where $\epsilon_{w}(k,n)$ is given by (\ref{eq:errorcc}) and (\ref{eq:errorir}) for Chase combining and incremental redundancy, respectively. The superscript is omitted for the simplicity of presentation.

\begin{figure}[t]
\centering
\includegraphics[width=\columnwidth]{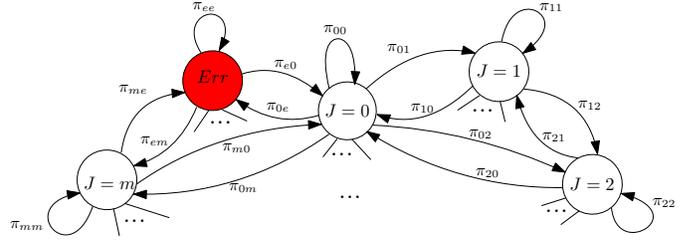}
\caption{The state diagram of the proposed D-HARQ scheme. The transmitter at state $J=j$, for $j\in[0,m]$, can send at most $L+j$ copies of a packet and state $J=e$ is the error state. }
\label{fig:MClimit}
\end{figure}
\subsection{Reliability and Throughput Analysis of D-HARQ}
We use a Markov model to analyze the dynamic of the proposed scheme. Fig \ref{fig:MClimit} shows the Markov model of the proposed scheme where each state, denoted by $J$, represents the number of additional re-transmissions. That is when the system is at state $J=j$, the transmitter can send at most $L+j$ copies of the packet. As we have limited the maximum number of additional re-transmissions, we have only $m+2$ states, including the error state. The system transits from state $J=i$ to the error state, when the receiver collected $L+i$ copies of the packet, through either CC or IR, and the decoding failed.

Let $\Pi$ denotes the state transition matrix for D-HARQ with parameters $L$ and $m$, then it can be shown as follows:
\[\Pi=\left[\begin{array}{cccc}
\pi_{00}&\cdots&\pi_{0m}&\pi_{0e}\\
\vdots&\cdots&\vdots&\vdots\\
\pi_{m0}&\cdots&\pi_{mm}&\pi_{me}\\
\pi_{e0}&\cdots&\pi_{em}&\pi_{ee}\end{array}\right],\]
where $\pi_{ij}$ is the probability of transiting from state $J=i$ to state $J=j$, for $i,j\in[0,m]$, and $\pi_{ie}$ (or $\pi_{ei}$) is the probability of transiting from state $J=i$ (or error state) to the error state (or $J=i$).

The following lemma characterizes the transition probability matrix of the proposed D-HARQ.
\begin{lemma}
In D-HARQ, where the interval between the deadlines of two consecutive packets is $L$ time slots and the packets can be transmitted no more than $L+m$ times, the transition probability $\pi_{ij}$, for $i,j\in [0,m]$ is given by
\[\pi_{ij}\approx\left\{\begin{array}{ll}
\epsilon_{L-j+i-1}(k,n)-\epsilon_{L-j+i}(k,n),& 0 \le j<m,\\
1-\epsilon_{L+i-m}(k,n),& j=m,\\
0,&\mathrm{o.w.,}\end{array}\right.\]
$\pi_{ie}\approx\epsilon_{L+i}(k,n)$, and 
\begin{align}
    \pi_{ei}\approx\epsilon_{L-i-1}(k,n)-\epsilon_{L-i}(k,n).
\end{align}
\end{lemma}
\begin{IEEEproof}
When the system is at state $J=i$, the transmitter can send at most $L+i$ copies of the packet. The system goes to state $J=j$, when the packet can be correctly decoded with exactly $L+i-j$ transmissions. This happens when the decoding failed with $L+i-j-1$ packets and succeeded with $L+i-j$ packets. We then have
\begin{align}
    \nonumber \pi_{ij}&=\mathrm{Prob}\left\{\bigcap_{w=1}^{L+i-j-1}\mathcal{A}_{w}\bigcap\mathcal{A}^c_{L+i-j}\right\}\\
    \nonumber &\overset{(a)}{\approx}\mathrm{Prob}\{\mathcal{A}_{L+i-j-1}\}-\mathrm{Prob}\{\mathcal{A}_{L+i-j}\}\\
 &=\epsilon_{L-j+i-1}(k,n)-\epsilon_{L-j+i}(k,n),
 \label{eq:error}
\end{align}
where step $(a)$ follows from $\mathrm{Prob}\{\mathcal{A}_w\bigcap\mathcal{A}_{w-1}\}=\mathrm{Prob}\{\mathcal{A}_w\}\mathrm{Prob}\{\mathcal{A}_{w-1}|\mathcal{A}_{w}\}\approx\mathrm{Prob}\{\mathcal{A}_w\}$, which is due to the fact that if the decoding failed with $w$ packets, it almost certainly failed with $w-1$ previous packets, i.e., $\mathrm{Prob}\{\mathcal{A}_{w-1}|\mathcal{A}_{w}\}\approx1$. This means to declare an error after $L+i-j$ transmissions, all previous transmissions should declare failure too.

The system goes to state $J=m$, when the packet can be decoded with $w$ transmissions where $0< w\le L+i-m$. Therefore we have 
\begin{align}
\nonumber \pi_{im}&=\mathrm{Prob}\left\{\overset{L+i-m}{\underset{w=1}{\bigcup}} \mathcal{A}^{c}_{w}\right\}=1-\mathrm{Prob}\left\{\overset{L+i-m}{\underset{w=1}{\bigcap}} \mathcal{A}_{w}\right\}\\
\nonumber &\overset{(b)}{\approx}1-\mathrm{Prob}\left\{ \mathcal{A}_{L+i-m}\right\}=1-\epsilon_{L+i-m}(k,n),
\end{align}
where step $(b)$ follows from the fact $\mathrm{Prob}\{\mathcal{A}_w\bigcap\mathcal{A}_{w-1}\}=\mathrm{Prob}\{\mathcal{A}_w\}\mathrm{Prob}\{\mathcal{A}_{w-1}|\mathcal{A}_{w}\}\approx\mathrm{Prob}\{\mathcal{A}_w\}$. 
The probability of transiting from state $J=i$ to the error state is the probability that the decoding failed when the receiver collects $L+i$ packets, i.e.,  $\pi_{ie}=\mathrm{Prob}\{\mathcal{A}_{L+i}\}=\epsilon_{L+i}(k,n)$. 

When the system is at state $J=e$, it transits to state $j=i$ if the decoding succeeds with exactly $L-i$ transmissions, i.e., 
\begin{align}
    \nonumber \pi_{ei}&\approx\mathrm{Prob}\{\mathcal{A}_{L-i-1}\}-\mathrm{Prob}\{\mathcal{A}_{L-i}\}\\
    \nonumber &=\epsilon_{L-i-1}(k,n)-\epsilon_{L-i}(k,n).
\end{align}
This completes the proof.
\end{IEEEproof}

We then calculate the packet error rate. In D-HARQ, packet error rate denoted by $\zeta_\mathrm{D}$ is the probability that the system is at the error state. The following lemma characterizes $\zeta_\mathrm{D}$. 
\begin{lemma}
Let $\mathrm{P_{stat}}=[p_0,p_1,\cdots,p_{m+1}]$ denotes the stationary distribution corresponding the transition matrix $\Pi$. The packet error rate of the proposed D-HARQ scheme with parameters $L$ and $m$, is then $\zeta_\mathrm{D}(k,n,L,m)=p_{m+1}$.
\end{lemma}
\begin{IEEEproof}
The lemma follows directly from the fact that the stationary distribution of the system can be characterized by the eigenvector of matrix $\Pi'$ corresponds to eigenvalue 1 and the packet error probability is simply the stationary probability of being at state $J=e$.
\end{IEEEproof}



\begin{remark}
The packet error rate for the proposed D-HARQ scheme with parameters $L$ and $m=1$ is given by
\begin{align}
    \nonumber&\zeta_\mathrm{D}(k,n,L,1)\\&\approx \frac{\epsilon_{L+1}(k,n)-\epsilon_{L-1}(k,n)\epsilon_{L+1}(k,n)+\epsilon_{L}^2(k,n)}{1-\epsilon_{L-1}(k,n)+\epsilon_{L}(k,n)},
\end{align}
which can be easily derived by solving the linear equation $(\Pi-I_3)\mathrm{P'_{stat}}=\mathbf{0}$, where $I_3$ is the $3\times 3$ identity matrix.
\end{remark}
\begin{lemma}
The throughput of the proposed D-HARQ with parameters $L$ and $m$ is given by
\begin{align}
    \eta_{D}(k,n,L,m)\approx\frac{{k}(1-\zeta_D(k,n,L,m))}{2n~\mathrm{P}'_{\mathrm{stat}}\left(\Pi\circ\Lambda\right)\mathbf{1}_{m+2}},
    \label{eq:ThD}
\end{align}
where $\mathbf{1}_{m+2}$ is the all-one column vector of length $m+2$, $\Pi\circ\Lambda$ is the Hadamard product of matrices $\Pi$ and $\Lambda$ and
\[\Lambda=\left[\begin{array}{ccccc}
L&L-1&\cdots&L-m&L\\
L+1&L&\cdots&L-m+1&L+1\\
\vdots&\vdots&\cdots&\vdots&\vdots\\
L+m&L+m-1&\cdots&L&L+m\\
L&L-1&\cdots&L-m&L\end{array}\right].\]
\end{lemma}
\begin{IEEEproof}
When the system is at state $J=i$, then it transits to state $J=i$ with probability $\pi_{ij}$ and the number of transmissions will be $L+i-j$, which is the $j^{th}$ element of the $i^{th}$ row of Matrix $\Lambda$. The transmitter will send $\pi_{ie}(L+i)+\sum_{j=0}^{m}\pi_{ij}(L+i-j)$ packets on average when it is at state $J=i$. This is equivalent to the $i^{th}$ row of $\left(\Pi\circ\Lambda\right)\mathbf{1}_{m+2}$. When the system is at the error state, it sends on average $\sum_{i=0}^{m}\pi_{ei}(L-i)$ packets. This is equivalent to the to the last row of $\left(\Pi\circ\Lambda\right)\mathbf{1}_{m+2}$. In the stationary state, the average number of transmissions can be then calculated by $\mathrm{P'_{stat}}(\Pi\circ\Lambda)\mathbf{1}_{m+2}$. As each packet transmission takes one mini slot of length $n$ symbols and requires a feedback from the receiver that takes place over a mini slot of length $n$ symbols, the throughput can be easily derived by (\ref{eq:ThD}). 
\end{IEEEproof}
\subsection{Reliability and Throughput Analysis of the Conventional HARQ and Fixed Transmission}
For the Fixed transmission scheme with $L$ transmissions, the packet error rate denoted by $\zeta_{F}(k,n,L)$ is given by
\begin{align}
    \zeta_{F}(k,n,L)=\epsilon_L(k,2n),
\end{align}
where $k$ is the information block length and the time slot duration is $2n$. Accordingly, the throughput is given by:
\begin{align}
    \eta_{F}=\frac{k(1-\epsilon_L(k,2n))}{2nL}.
\end{align}

In the conventional HARQ with the maximum $L$ transmissions, the packet error probability, denoted by $\zeta_{H}(k,n,L)$ can be calculated as follows:
\begin{align}
    \zeta_{H}(k,n,L)=\epsilon_{L}(k,n),
\end{align}
due to the facts that a feedback message is sent by the receiver after each packet transmission, which consumes one mini-slot and the maximum number of transmissions is $L$. 

In conventional HARQ, the decoding will succeeds after receiving exactly $i$ packets $1\le i\le L$ packets with probability $\approx\epsilon_{i-1}(k,n)-\epsilon_i(k,n)$. This follows directly with the same argument as in (\ref{eq:error}). The decoding will fail and the packet will drop after $L$ transmissions with probability $\epsilon_L(k,n)$. Therefore, the average number of transmissions is $L\epsilon_L(k,n)+\sum_{i=1}^{L}i(\epsilon_{i-1}(k,n)-\epsilon_{i}(k,n))$. The throughput of the conventional HARQ scheme, denoted by $\eta_{H}(k,n,L)$, can be then calculated as follows:
\begin{align}
    \nonumber&\eta_{H}(k,n,L)\\
    &\approx\frac{k(1-\epsilon_L(k,n))}{2n\left(L\epsilon_{L}(k,n)+\sum_{i=1}^{L}i(\epsilon_{i-1}(k,n)-\epsilon_{i}(k,n))\right)}.
\end{align}

\section{Numerical Results}
We first consider a simple communication scenario with $L=2$ and $m=1$. This scenario is very important as in many URLLC applications, only one re-transmission is allowed due to the very strict delay constraint. We will show that the proposed scheme achieves higher level of reliability by effectively provide more opportunities for packet re-transmissions for URLLC applications. In the fixed transmission scheme exactly two copies of the packet of length $2n$ is sent to the receiver. However in D-HARQ, packets of length $n$ are sent to the receiver, leaving half of the timeslot for the ACK/NACK transmission. 

\begin{figure}[t]
\centering
\includegraphics[width=1\columnwidth]{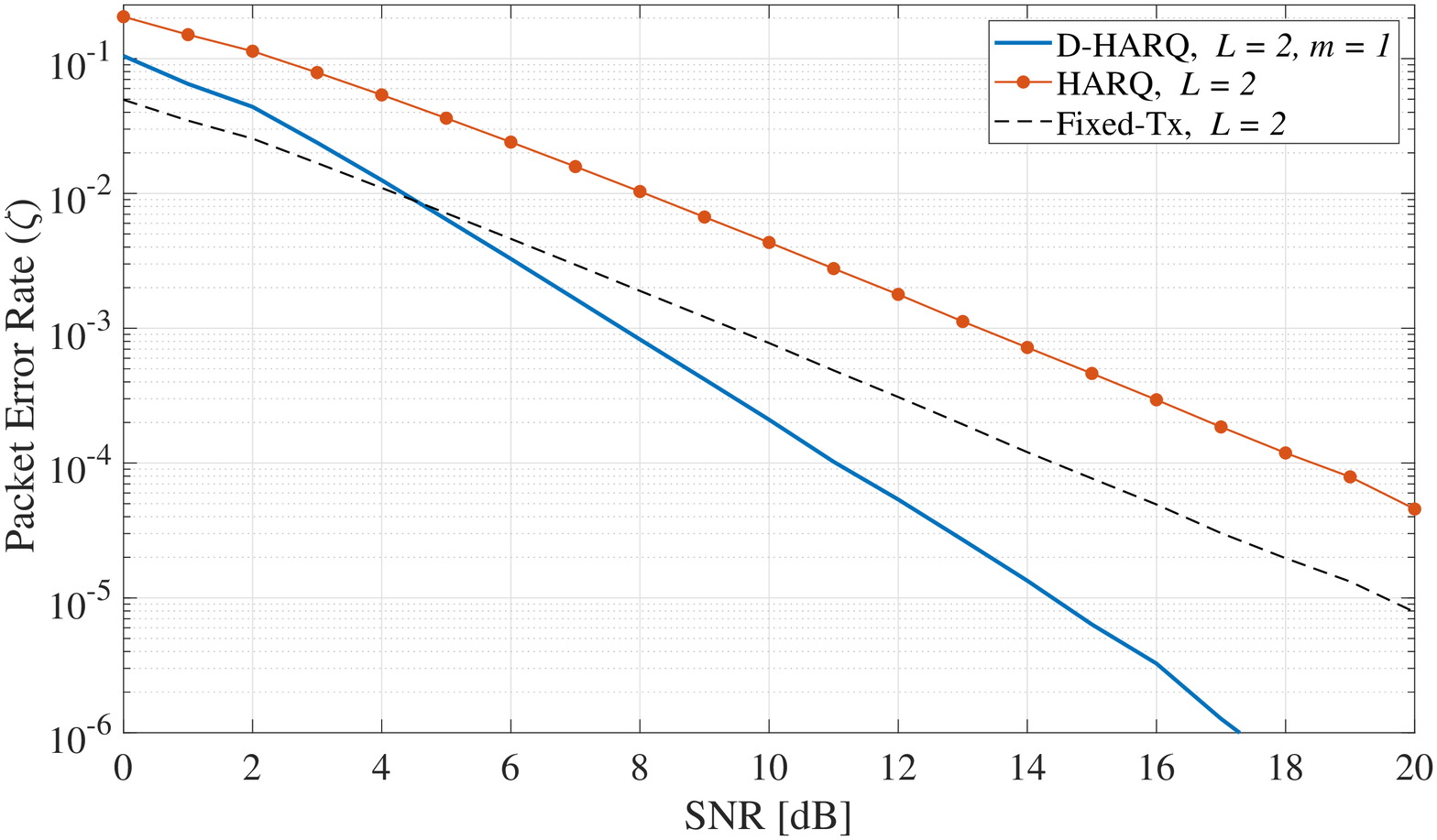}
\vspace{-4ex}
\caption{Packet error rate versus SNR when the time slot duration is 64 symbols and $k=32$. }
\label{fig:sim1}
\end{figure}

Fig. \ref{fig:sim1} shows the packet error rate (PER) versus SNR when $L=2$, $m=1$, $k=32$. As can be seen in this figure, the proposed D-HARQ significantly outperforms the fixed transmission scheme in terms of PER. More specifically, at the target PER of $10^{-4}$, D-HARQ achieves 3dB gain compared with the fixed transmission scheme. It is important to note that in Fixed-Tx, the transmitter can use a lower code rate to encode each packet compared with D-HARQ. In particular, when the feedback mini-slot is half of the timeslot duration, the channel code rate of the fixed transmission scheme is half that of the D-HARQ. Fixed-Tx outperforms conventional HARQ as in HARQ a shorter packet is used due to the feedback transmission.

\begin{figure}[t]
\centering
\includegraphics[width=1\columnwidth]{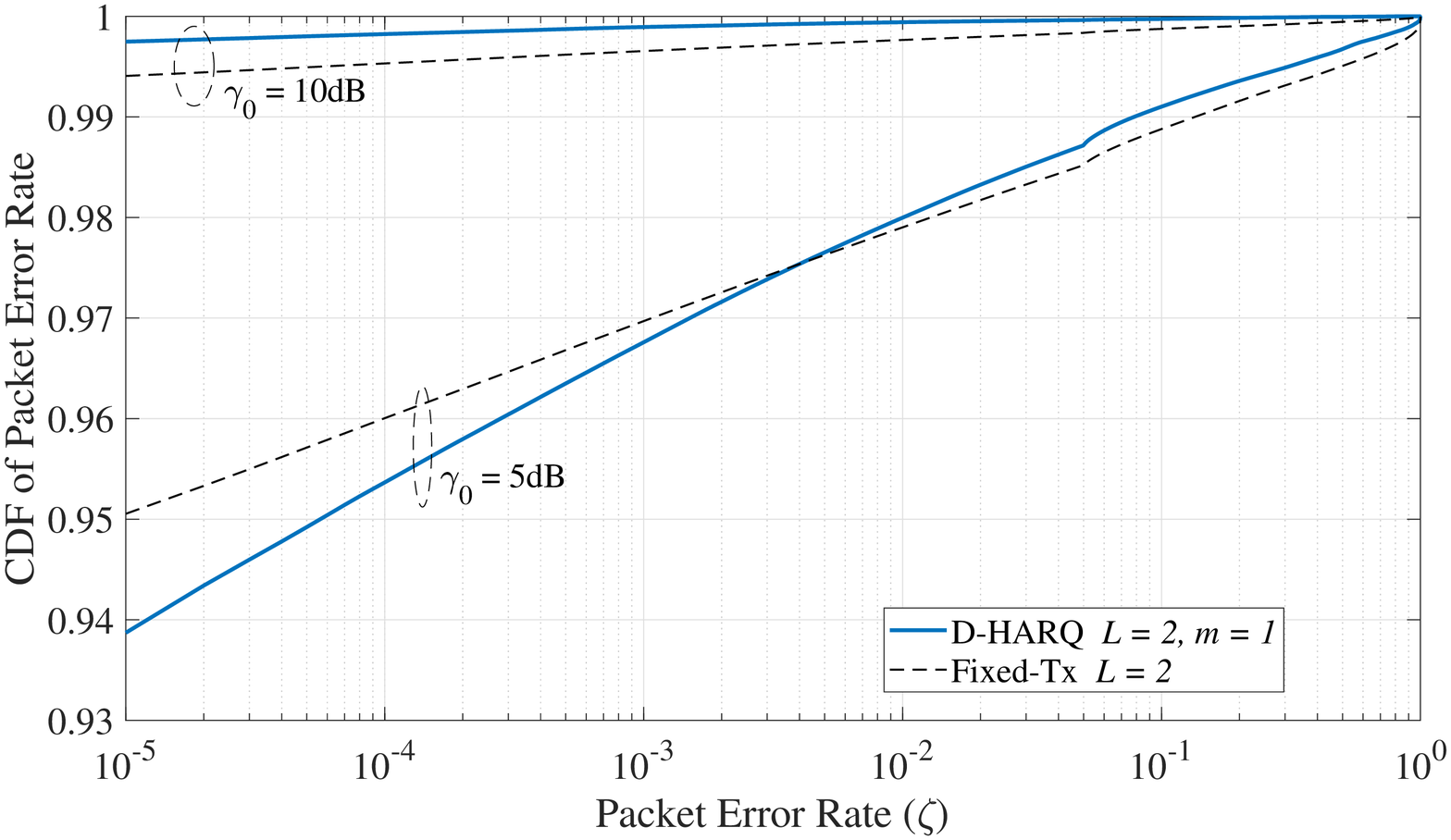}
\vspace{-4ex}
\caption{The CDF of the packet error rate at different SNRs when the time slot duration is 64 symbols, $k=32$, $L=2$ and $m=1$.}
\label{fig:simCDF}
\end{figure}

Fig. \ref{fig:simCDF} shows the cumulative distribution function (CDF) of the packet error rate. As can be seen in this figure, D-HARQ achieves a much lower packet error rate compared with the fixed transmission scheme. This is of particular importance for URLLC applications as every packet must be received with the desired level of reliability by the target deadline. As can be seen in Fig. \ref{fig:simCDF}, at 10 dB, D-HARQ achieves a very high reliability with a large probability compared with the fixed transmission scheme. 

\begin{figure}[t]
\centering
\includegraphics[width=1\columnwidth]{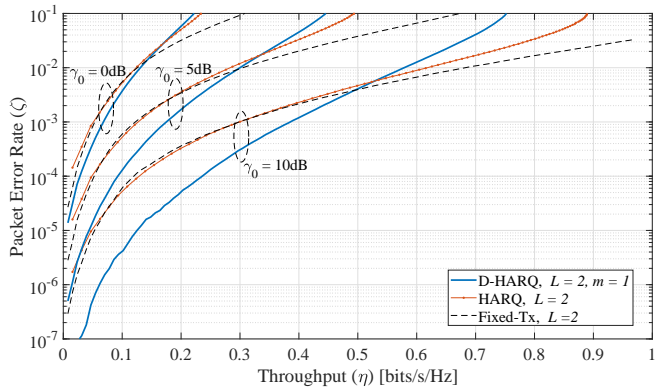}
\vspace{-4ex}
\caption{Packet error rate versus throughput at different SNRs when the timeslot duration is 64. }
\label{fig:PERvsTh}
\end{figure}

Fig. \ref{fig:PERvsTh} shows the packet error rate versus the throughput. One can use different channel code rates to encode each packet of data. This figure shows that the proposed D-HARQ achieves a much higher throughput at the same target PER. More specifically, at $\gamma_0=10$dB under the same delay constraint $L=2$ and target packet error rate $10^{-4}$, D-HARQ achieves almost double the throughput of Fixed-Tx. We can conclude here that in D-HARQ even though we are sending shorter packets due to time reservation for feedback transmission, we can get a higher throughput compared with the fixed transmission scheme under the same reliability constraint or achieves a much lower packet error rate under the same throughput thanks to the potentially larger number of diversity branches in the proposed D-HARQ scheme. 

\begin{figure}[t]
\centering
\includegraphics[width=1\columnwidth]{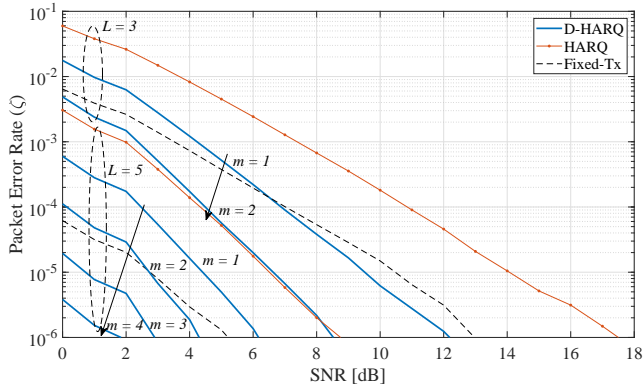}
\vspace{-4ex}
\caption{Packet error rate versus SNR when the timeslot duration is 64 symbols. }
\label{fig:sim4}
\end{figure}

Next, we show the performance of the proposed D-HARQ when the delay constraint $L$ increases. As can be seen in Fig. \ref{fig:sim4}, by increasing $m$ the proposed D-HARQ achieves a much lower error probability compared with the Fixed-Tx and conventional HARQ schemes. This is of particular importance for URLLC applications that D-HARQ achieves a higher level of reliability without imposing a delay into the system. In fact in D-HARQ, we are using the available diversity branches in a more effective way that is the packets can be opportunistically transmitted more times when less diversity branches were used in the previous round of packet transmission.

\section{Concluding Remarks}
In this paper, we proposed a dynamic HARQ (D-HARQ) scheme for ultra-reliable delay sensitive communications. In the proposed scheme, each packet can be re-transmitted more times if the previous packet was decoded earlier than its deadline. We analyzed the reliability and throughput of the proposed D-HARQ and showed that it is superior to the fixed transmission and conventional HARQ. The proposed scheme opportunistically provides more diversity without imposing any delay into the system. In particular, when each packet must be correctly received with a maximum two transmissions, the proposed D-HARQ achieves 10dB gain compared with conventional HARQ thanks to effectively using the diversity branches in a dynamic manner.

\bibliographystyle{IEEEtran}
\footnotesize
\bibliography{Ref}
\end{document}